# Synthesis of realistic fetal MRI with conditional Generative Adversarial Networks

Marina Fernandez Garcia[1], Rodrigo Gonzalez Laiz[1], Hui Ji[1], Kelly Payette[1], Andras Jakab[1,2]

1: Center for MR-Research, University Children's Hospital Zürich, Zürich, Switzerland
2: University of Zürich, Zürich, Switzerland

Corresponding author: András Jakab, Email: Andras.Jakab@kispi.uzh.ch

Fetal brain magnetic resonance imaging serves as an emerging modality for prenatal counseling and diagnosis in disorders affecting the brain. Machine learning based segmentation plays an important role in the quantification of brain development. However, a limiting factor is the lack of sufficiently large, labeled training data. Our study explored the application of SPADE, a conditional general adversarial network (cGAN), which learns the mapping from the label to the image space. The input to the network was super-resolution T2-weighted cerebral MRI data of 120 fetuses (gestational age range: 20 – 35 weeks, normal and pathological), which were annotated for 7 different tissue categories. SPADE networks were trained on 256 x 256 2D slices of the reconstructed volumes (image and label pairs) in each orthogonal orientation. To combine the generated volumes from each orientation into one image, a simple mean of the outputs of the three networks was taken. Based on the label maps only, we synthesized highly realistic images. However, some finer details, like small vessels were not synthesized. A structural similarity index (SSIM) of 0.972 ± 0.016 and correlation coefficient of 0.974 ± 0.008 were achieved. To demonstrate the capacity of the cGAN to create new anatomical variants, we artificially dilated the ventricles in the segmentation map and created synthetic MRI of different degrees of fetal hydrocephalus. cGANs, such as the SPADE algorithm, allow the generation of hypothetically unseen scenarios and anatomical configurations in the label space, which data in turn can be utilized for training various machine learning algorithms. In the future, this algorithm would be used for generating large, synthetic datasets representing fetal brain development. These datasets would potentially improve the performance of currently available segmentation networks.

## Background

According to the WHO, one of the leading causes of child mortality worldwide is congenital diseases [1]. Fetal magnetic resonance imaging (MRI) scans are becoming more and more important to examine the development of the fetus, as well as to identify diseases like spina bifida, intrauterine growth retardation and congenital heart disease [2, 3]. These scans can also help us understand neurodevelopment in both pathological and non-pathological cases. However, the analysis of such MRI data requires specialized clinical expertise, and quantitative analysis by experts is prone to human errors and observer bias. Furthermore, such analysis remains time consuming.

One important goal of the quantitative analysis of fetal MRI data is to characterize prenatal brain development and link the morphological impairments of the developing brain to clinical outcomes or identify certain pathologies. This can be achieved by segmenting the MRIs at different gestational ages into different cerebral tissue compartments and anatomical structures of the developing brain. As manual segmentation is costly, time consuming and leads to inter-rater variability, automated methods would offer advantage. An important bottleneck in such endeavours is the availability of training data and expert annotations, which data synthesis approaches would potentially overcome.

Various segmentation networks, typically U-nets [4], were trained with manually segmented MRIs for the FeTA challenge 2021 [3], but there was still room for improvement, since the machine-learning based methods appeared to reach a plateau of performance in the mid-high range of accuracy. Furthermore, data is often scarce and manual annotations are time-consuming.

A common approach for improving the performance of a network is data augmentation. There are various methods for this, but in general, they can be classified into three different types of data augmentation methods: basic, deformation based, and deep learning based methods. Basic augmentation methods include geometric transformations (e.g. rotations, translations, flipping and scaling), cropping, intensity operations and noise injection [5]. Some of these methods were already incorporated into the U-net training and thus would probably not enhance the networks performance drastically. Deformation based methods are used to make the model more generalisable by applying non-linear deformations within user defined limits. This is done, for example, through random displacement fields or interpolation between different cases.

Finally, the last category, deep learning based augmentation methods, was the topic of discussion in this case. This mainly includes variational autoencoders (VAEs) and Generative Adversarial Networks (GANs). The purpose of our work was to synthesize realistic fetal MRI data from label maps only with the aim to provide new synthetic data for training machine learning models for fetal MRI analysis.

## Methods

*Dataset.* T2-weighted single shot fast spin echo images of 120 fetuses (gestational age range: 20 – 35 weeks, normal and pathological) were acquired in three orthogonal planes, acquired on a 1.5T or 3T MRI. These scans were motion corrected and reconstructed into a super-resolution volume (SR) for each subject. Each volume was manually annotated into seven cerebral tissues. 80 datasets were used for training SPADE and 40 for testing. Further description of the fetal MRI processing and the dataset is found in [3] and we illustrated one MRI dataset with annotations in **Figure 1**.

*Generative adversarial networks.* GANs were originally developed by I. Goodfellow [6] and consist of two main components: a generator and a discriminator (**Figure 2**). The generator is the network you are trying to train such that it learns to produce new, fake images, similar to the ones already in your data-set. The discriminator is a network that is being fed both real and fake (synthesized by the generator) images, and its aim is to correctly distinguish the real from the fake images. The method behind a GAN can be thought of as a 'competition' between these two components, where the generator is trying create images realistic enough to trick the discriminator into thinking they are real, whilst the discriminator is trying to correctly classify each image it is given into real and fake.

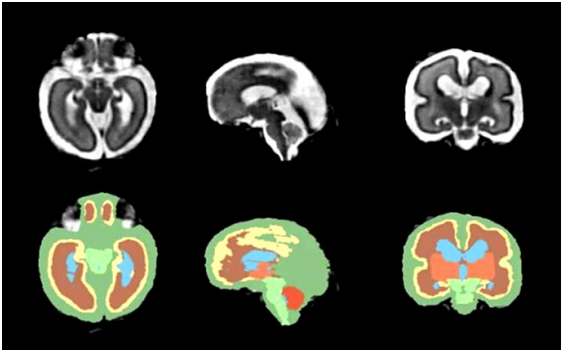

*Figure 1. Example of an input fetal MRI dataset. Top row: T2-weighted super-resolution MRI. Bottom row: manual annotations into seven tissue categories: external fluid, gray matter, white matter, deep gray matter, ventricles, cerebellum and brainstem.*

The two networks are simultaneously being trained by trying to minimise what is called a mini-max loss function [8]:

$$L(x, z) = E_x[log(D(x))] + E_z[log(1 - D(G(z)))] \quad (1)$$

Where:
$D(x)$ = the discriminator's estimate of the probability that real data instance x is real.
$E_x$ = the expected value over all real data instances.
$G(z)$ = the generator's output when given noise z.
$D(G(z))$ = the discriminator's estimate of the probability that a fake instance is real.
$E_z$ = the expected value over all random inputs to the generator.

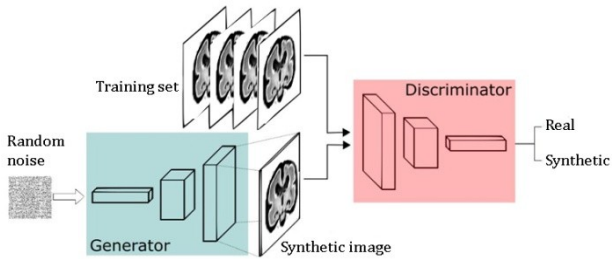

*Figure 2. General structure of a GAN*

Ideally, if the settings are chosen correctly, towards the end of the training the generator and discriminator loss should both converge to 0.5. This would indicate that about half of the images created by the generator are determined to be real, since the discriminator can no longer tell which images are real and which ones are fake.

SPADE [9], also referred to as GauGAN, is a type of conditional GAN that learns to map from the label space to the image space. Instead of only looking at images, SPADE looks at label and image pairs and learns to create realistic images based on the given label map. Having this same application available to map from fetal brain label maps to realistic seeming MRIs would be useful for a number of reasons. First, one could create an indefinite amount of data from already existing labels. This data could then be used to retrain the U-net to attempt to improve its performance. Moreover, one could also create new labels, for example interpolating between different gestational ages, to analyse the fetal development in an MRI format.

Given that SPADE is a network developed for 2D images, an MRI orientation had to be chosen to train the network each time. Since the MRIs were manually segmented in the axial orientation, in the other orientations there might be some inconsistencies between neighbouring slices. Thus, it made sense to train the network in various orientations and then combine these to regain a 3D structure. The SPADE network was trained once for the sagittal, axial and coronal orientation.

Two different SR reconstruction methods were used to create the MRIs. To see the effect of the SPADE performance for the two methods, the three orientations were also trained separately for the MRI with the total variance approach reconstruction (labelled as 'mial'), as well as the irtk-simple method ('irtk').

***Evaluation of the GAN performance.*** To examine the performance of the SPADE network, various scores were considered that assessed the similarity between the original and synthesized images. One common way to evaluate image similarity is by measuring the mean squared error (MSE) between the two. Here, the absolute difference between the intensity of each pixel is measured and added together. However, since we want, to an extent, a change in intensity (to not obtain the exact same MRI), this was not deemed a good performance metric. Rather, we are looking for a value that assesses the structural similarity, and not the absolute one.

Thus, we arrive at the structural similarity index measure (SSIM), which is a measure of the perceived similarity between two images, which, different to MSE, looks at the structural similarity between two images [10]. This is much closer to the way the human eye perceives similarities than MSE. SSIM takes into account three factors: luminance, contrast and structure.

Moreover, mutual information (MI) was also considered as an evaluation metric. Mutual information is a measure of the dependence between two variables, or in this case two images. It essentially tells you how much information is contained in the first image about the later one [11].

Another way to measure dependence between two images are correlation coefficients, which are a measure of the statistical relationship between two variables [12]. The correlation coefficient ranges from -1 to 1, getting closer to 1 when there is a very strong positive correlation between the two variables.

Finally, the fréchet inception distance (FID) is a metric used to assess the quality of images generated specifically by a neural network, for example of a GAN. It compares the distribution of generated with images, with that of the training set [13]. The FID score is calculated as follows:

$$FID = ||\mu - \mu_w||_2^2 + tr(\Sigma + \Sigma_w - 2(\Sigma^{1/2}\Sigma_w\Sigma^{1/2})^{1/2}) \quad (2)$$

Here $\mu$ and $\mu_w$, and $\Sigma$ and $\Sigma_w$, are the mean and covariances of the original and synthesized distributions respectively. Whilst this metric agreed with our evaluation of the different models, it doesn't give you a score per MRI, and only gives you one score for a whole data set instead of a distribution of scores. It is also dependent on the number of images used to calculate the score, so it didn't seem like a useful metric to compare models that had been trained with and tested on data sets of different sizes.

After comparing various image pairs with their given scores, SSIM and the correlation coefficient were determined as the ones that best represented the score that the human eye would give. MI did not seem to align a lot with what we perceived as good performance, so this was no longer considered. Since SSIM seemed slightly more accurate, a self defined score was developed, combining the two metrics:

$$\text{Self defined score} = \frac{1}{3} * (2 * SSIM + \text{Correlation coefficent})$$

***Hyperparameters.*** One of the parameters considered for the training of SPADE was the number of epochs. To assess the ideal number of epochs, the SPADE was trained with a smaller data set (only one slice from each MRI in the training set, instead of all), for different number of epochs. The SPADE training network, with the options "niter" and "niter decay" (see SPADE training guide), allows you to choose how many epochs you want with constant and how many with a gradually decreasing learning rate, respectively.

Here, the discussed metrics were compared for 25 epochs with constant learning rate, followed by 25 with a decaying learning rate, 50 epochs with constant learning rate, and 100 epochs with constant learning rate. The trained networks were then applied to unseen labels from the testing set, and their performance was measured by looking at the MSE, SSIM, Correlation Coefficient and self defined scores. As can be seen in **Figure 3,** 50 epochs clearly showed the best performance, indicating that 100 epochs were not necessary and might even lead to over-fitting. The scores also suggest that the combination of 25 constant and 25 decaying learning rate epochs were maybe not sufficient to reach a good network.

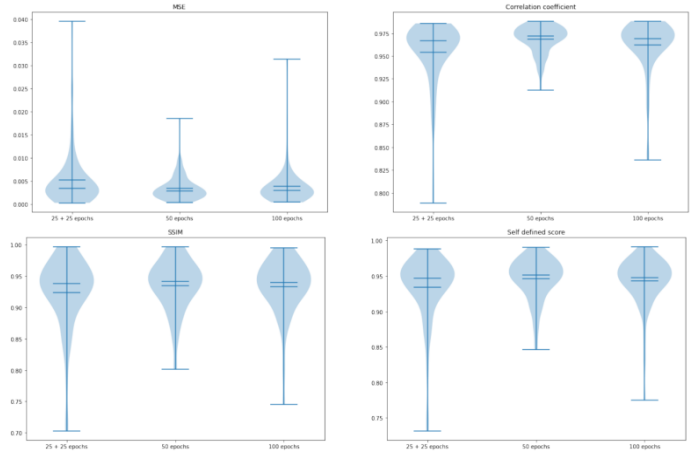

***Figure 3:*** *Score distributions for different number of epochs*

Furthermore, various learning rates were also explored, also training with the smaller training data-set. This included learning rates of 5E-5, 1E-4, 5E-4 and 1E-3. Since the learning rates showed quite similar performance (see figure A.1), maybe slightly better for around 1E-4, the default learning rate of 2E-4 was decided upon. The default batch size of size 1 was taken, although a larger batch size is a possible alternative that could be explored in the future.

## Results

***Performance of SPADE on fetal MRI data.*** In **Figure 4,** the performance of randomly chosen slices in the, from top to bottom, sagittal, coronal and axial orientation can be seen. Whilst the images look fairly similar, the main thing to look for, to analyse the network's performance, is that the structure of the labels remains intact and the difference in intensities between the different classes remains approximately constant. Since this is relatively difficult to judge visually, **Table 1** shows the MSE, SSIM, correlation coefficient and self defined scores between the original (middle) and synthesized (right) images. See figures A.2 and A.3 in the appendix to see some examples of the networks trained on only the 'irtk' and 'mial' networks applied to labels in each orientation.

|  | MSE | SSIM | CORR | Self defined |
|---|---|---|---|---|
| x (sagittal) | 0.00 | 0.97 | 0.98 | 0.97 |
| y (coronal) | 0.00 | 0.95 | 0.98 | 0.96 |
| z (axial) | 0.00 | 0.93 | 0.96 | 0.94 |

***Table 1.*** *Calculated scores for the example images in Figure 4.*

Given that the network was trained separately for three orientations, a way to combine them had to be decided upon. One way was to simply take a simple mean of the three intensity values for each voxel. This seems to work quite well, although it has a considerable blurring effect and doesn't allow you to consider outliers. Another method that was considered was to, for each voxel, out of the three values, take the two values closest to each other and average them. However, this was not achieved in a non-vectorized way, so it was rather time consuming and was therefore not chosen as an apt method.

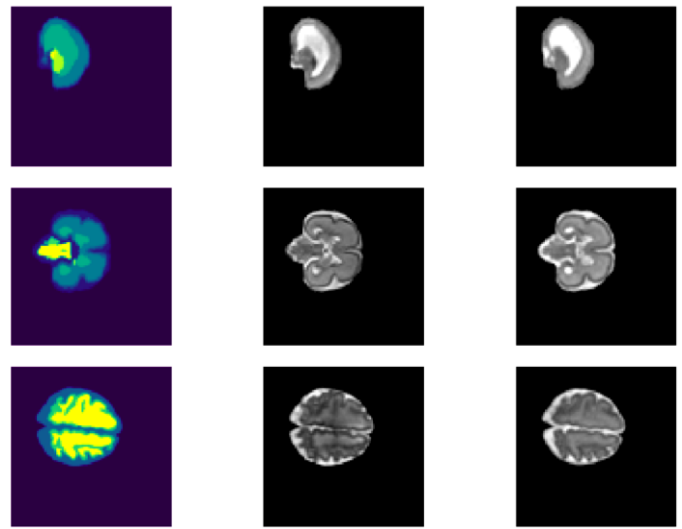

***Figure 3:*** *Example of each of the SPADE networks trained on each orientation for all the training data. On the left you can see the input label, in the middle the original MRI scan, and on the right the synthesized image.*

Finally, an outlier detection method was considered, that allowed you to choose how far from the median deviation your values should be. That way, a value 'm' could be chosen, based on how permissive you want to be with what is considered an outlier. It was, however, observed that this method sometimes led to some artifacts in the MRIs, which could stem from cases where the values that were closer together were in fact outliers. The combined MRI is obtained as follows (**Listing 1**):

```python
import numpy as np

def reject_outliers(data, m):   #a larger m means less outliers
    assert m > 0 , "Pick a positive number for the comb_mode parameter."
    epsilon = 0.0001
    median = np.median(data, axis = 3)
    median3 = np.stack([median,median,median], axis = 3)
    d = np.abs(data - median3)
    mdev = np.median(d, axis = 3) + epsilon
    mdev3 = np.stack([mdev,mdev,mdev], axis = 3)
    s = d/mdev3
    new_data = np.where(s< m, data, median3)
    return np.mean(new_data, axis= 3)
```

***Listing 1.*** *Outlier rejection function as defined on python. The data input is a stack of the mris obtained for each orientation.*

To compare the different combination methods, their scores were evaluated in 3D, in order to assess the coherence of synthesized MRIs between the different layers. **Figure 4.** shows the 3D MSE, SSIM, correlation coefficient and self-defined scores for the three orientations and the three discussed combination methods. It is clear that taking the mean of the three orientations allows us to obtain a better score than each of the individual orientations. When comparing the combination methods, there was not a large difference between their performances, although in general, the mean method showed lower variability of scores.

Similarly, **Figure 5.** shows the 3D scores for the different SR reconstruction methods. These scores were obtained by applying each network to the full validation set (40 MRIs), so also to labels that did not stem from the SR reconstruction method the networks were trained on. Interestingly, the 'irtk' network performs almost as well as the 'all' network, even though it was trained on a data-set that was about half the size. The slightly worse performance of the 'irtk' network could stem from the fact that this SR reconstruction method has a slightly lower resolution, meaning that there is probably also a larger variance in the manual segmentation labels.

***Synthesis from manipulated label maps: case of ventriculomegaly.*** As previously mentioned, a benefit of SPADE is that one can artificially create a hypothetical scenario by altering the label map and seeing how this would look in the MRI format. We looked at the effect of artificially dilating the label map values corresponding to the ventricles to see if the trained SPADE network would come up with an accurate representation of the pathological scenario. In **Figure 6**, the different degrees of hydrocephalus can be seen, in sagittal orientation, obtained from the label map of a non-pathological case (see figures A.4 and A.5 in the appendix for the coronal and axial orientations). The top row shows the original label and MRI in the two leftmost columns. The middle column shows the synthesized MRI, based on the network trained on all training data. The next two columns show the synthesized MRIs for the irtk and mial trained datasets. In the following rows, different grades of hydrocephalus can be seen, where the ventricle labels were gradually dilated. The 'original' column in this case shows the transformation of the initial MRI if the same transformation is applied to the MRI as to the label map. Based on a visual analysis, SPADE seems to interpret the different grades of hydrocephalus well.

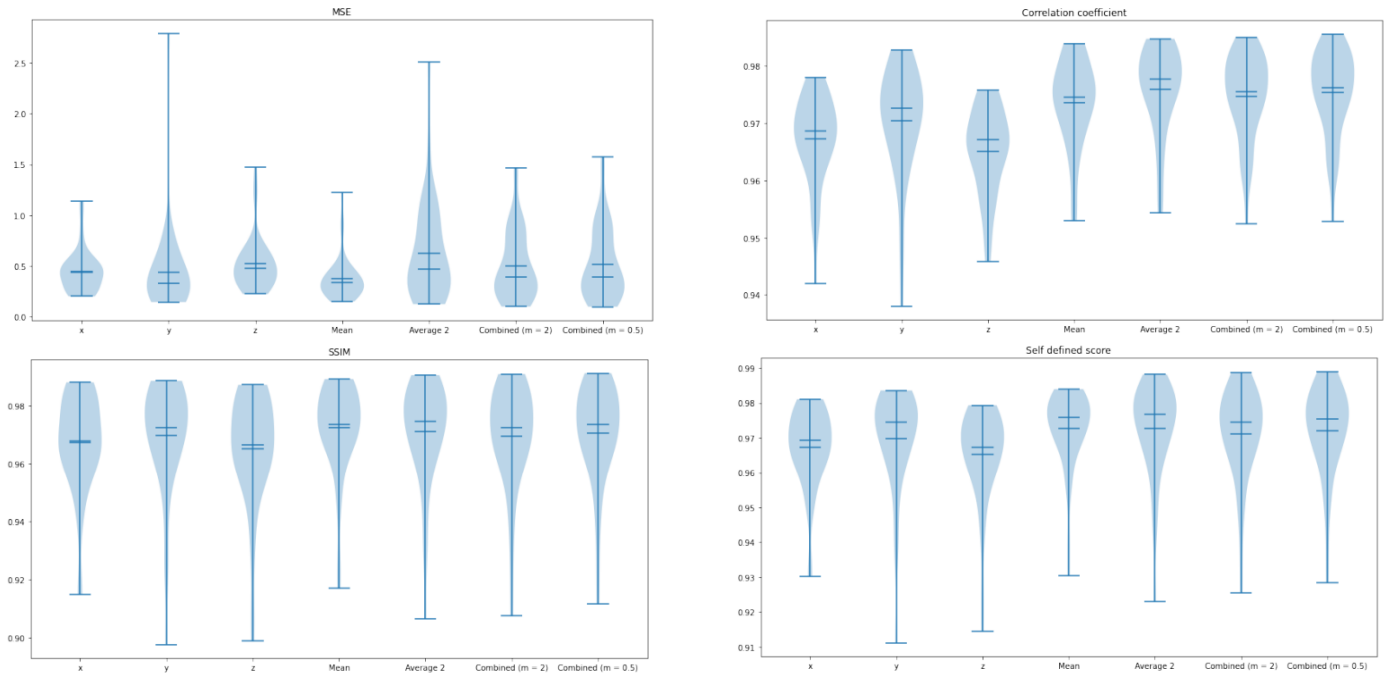

*Figure 4: Score distributions for different orientations and combination methods. x, y and z represent the scores of the networks only trained on the sagittal, coronal and axial orientation, respectively. 'Average 2' combines the three orientations by taking the two closest values for each voxel and taking their average. 'Combined (m = 0.5)' and 'Combined (m = 2)' show the scores of the outlier rejection method, where the parameter m (see Listing 1) is set to 0.5 and 2, respectively.*

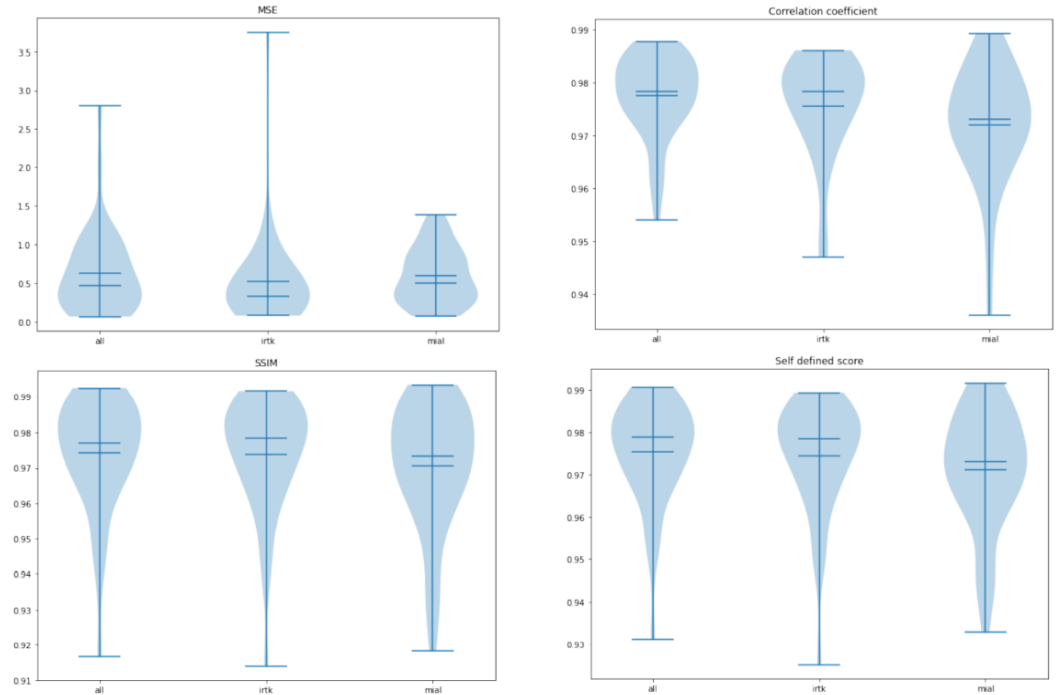

*Figure 5: Score distributions for different SR reconstruction methods. 'all', 'irtk' and 'mial' represent the scores of the networks where all three orientations were trained on either: the full data-set, only 'irtk' reconstructed mris and only 'mial' reconstructed mris, respectively. Here the three orientations for each method were combined taking the mean.*

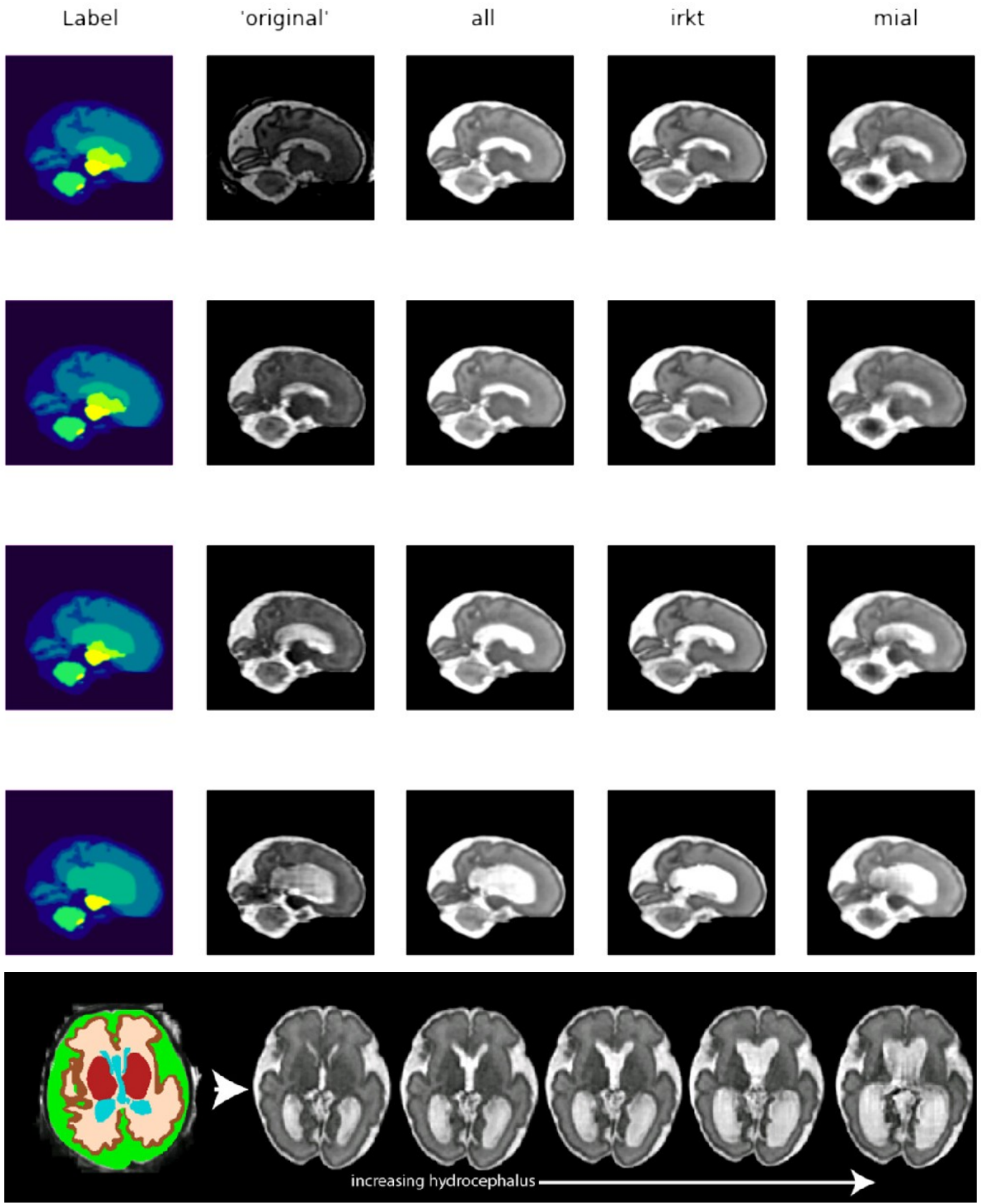

**Figure 6.** *Artificially created scenario with varying degrees of hydrocephalus, which was achieved by dilating the label corresponding to the lateral cerebral ventricles. Sagittal orientation. The original MRI vs. the synthetic images with different reconstructions are shown on the right.*

## Discussion

In this work, we have shown how a cGAN can be used to synthesize fetal MRI data using label annotations only. In principle, the method can be expanded to synthesize images that resemble any specific MR reconstruction methods (in our example: 'mials' or 'irtk'), or can synthesize new MRI data after manipulating (e.g. morphological operations) the label images.

An apparent limitation of our work is that one needs an anatomically valid label map with exactly the same type of classes as in our analysis. The data generated by the SPADE network is not entirely 'new' since it is based upon already existing labels, meaning that maybe the U-net does not gain all that much information from these image and label pairs. To ensure that the data is entirely new, one would also require the synthesis of new label maps. This was attempted as part of the project using a DCGAN, but the results were not very convincing as the network would gravitate towards a certain type of MRI slice, which is a common problem when training GANs. Therefore, we decided not to include the DCGAN based analysis in our current report. In addition, a GAN, or in this case DCGAN, is not thought for synthesizing images with discrete classes, so discretizing the output of the GAN such that it actually represents a label map is also something that would have to be tackled.

Furthermore, a limitation of our work is that the data-set used to decide on the hyperparameters was too small. To have a better informed evaluation of the parameters, one should use a larger data-set and explore a larger range of epochs and learning rates. For example, for the number of epochs, one could also consider the case of 50 epochs of constant learning rate and 25 decaying. In addition, there are more parameters that could be assessed when training SPADE, such as the batch size, the number of generator and discriminator filters and maybe even the type of optimizer. Whilst the default settings for these parameters worked quite well, there is probably still room for improvement.

Our method allowed the synthesis of fetal MRI based on real and modified label maps. SPADE successfully synthesized highly realistic MRI of the fetal brain, where all tissue types were clearly distinguishable and non-annotated details, such as the choroid plexus, germinal matrix, and some flow artifacts in the ventricles were visualized. However, some finer details, like small vessels were not synthesized. Possible applications of such synthetic datasets would be data augmentation in the training of image segmentation networks, for networks that aim to discriminate pathological vs. non-pathological fetal MRI, or to estimate gestational age from fetal MRI data.

## Acknowledgements

The project was funded by the Prof. Max Cloetta Foundation, the EMDO Foundation, the Vontobel Foundation and the Novartis Foundation for Medical-Biological Research.

The pregnant women, parents or legal guardians of infants gave written informed consent for the further use of their health-related data in research. The ethical committee of the Canton of Zurich approved the retrospective studies that collected fetal and neonatal MRI data (KEK decision numbers: ID 2016-01019 and ID 2022-01157).

**This manuscript is the author's original version.**

# A Appendix

## A.1 Parameters

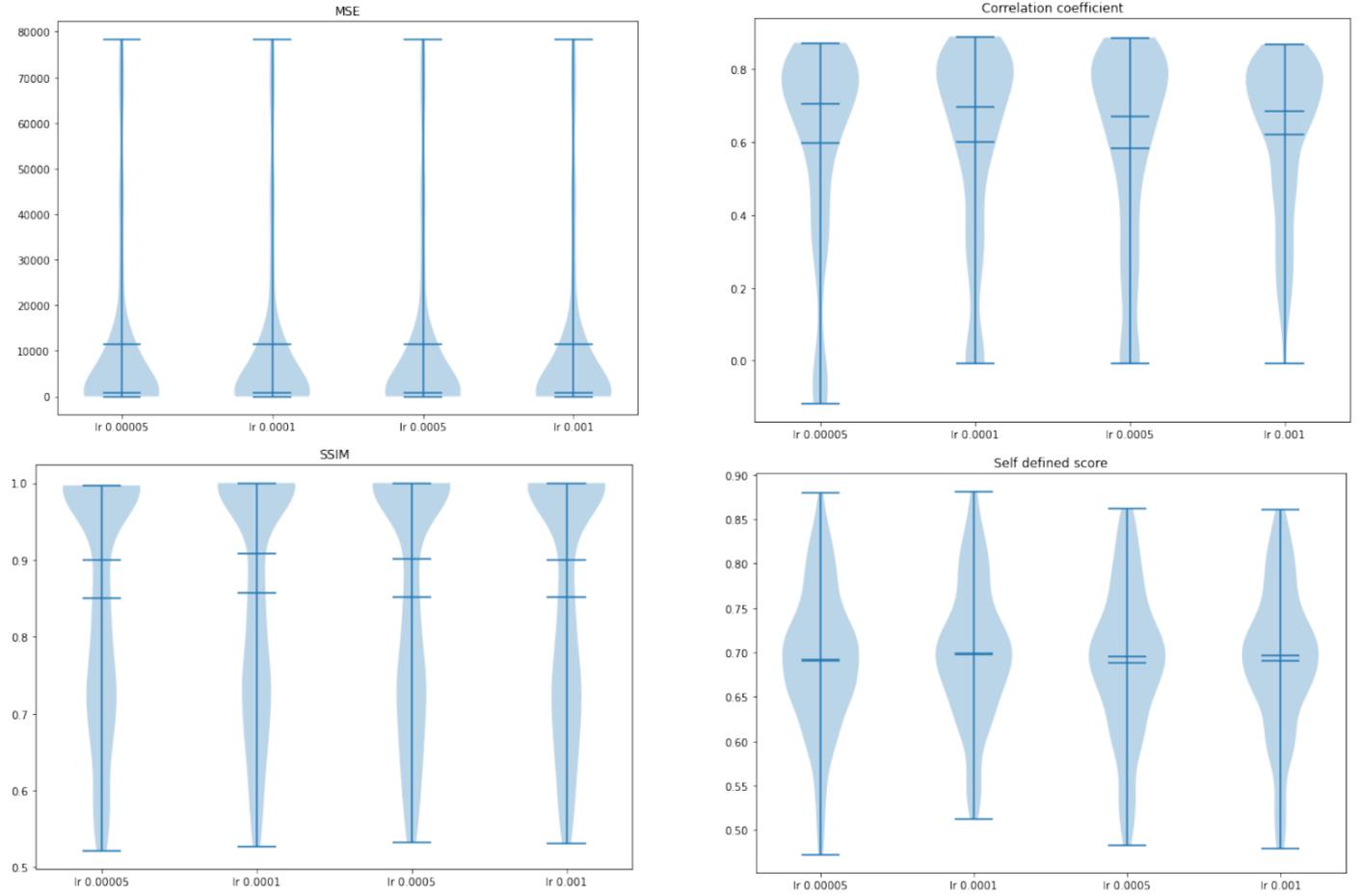

Figure A.1: Score distributions for different learning rates

## A.2 Irtk and mial examples

|  | MSE | SSIM | CORR | Self defined |
|---|---|---|---|---|
| x (sagittal) | 0.02 | 0.87 | 0.92 | 0.87 |
| y (coronal) | 0.00 | 0.95 | 0.98 | 0.96 |
| z (axial) | 0.01 | 0.91 | 0.97 | 0.93 |

Table A.1: Calculated scores for the example images trained with the irtk data in figure A.2

|  | MSE | SSIM | CORR | Self defined |
|---|---|---|---|---|
| x (sagittal) | 0.00 | 0.94 | 0.97 | 0.95 |
| y (coronal) | 0.00 | 0.93 | 0.97 | 0.95 |
| z (axial) | 0.00 | 0.94 | 0.97 | 0.95 |

Table A.2: Calculated scores for the example images trained with the 'mial' data in figure A.3

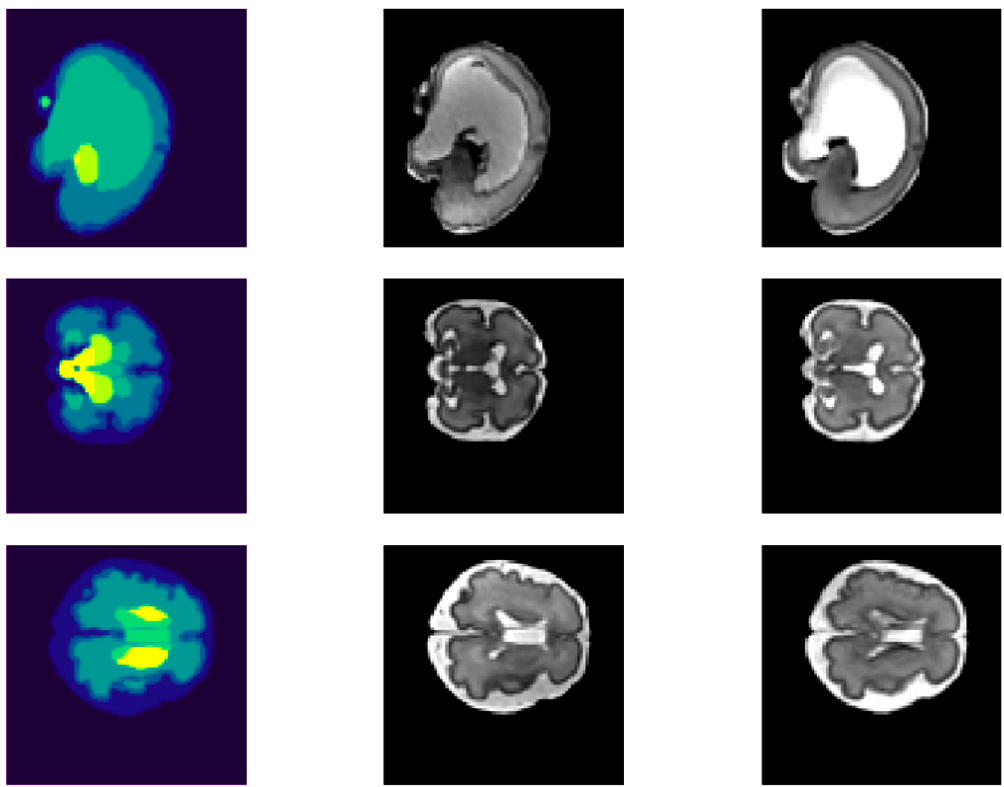

Figure A.2: Example of each of the SPADE networks trained on each orientation for only the 'irtk' the training data. On the left you can see the input label, in the middle the original MRI scan, and on the right the synthesized image.

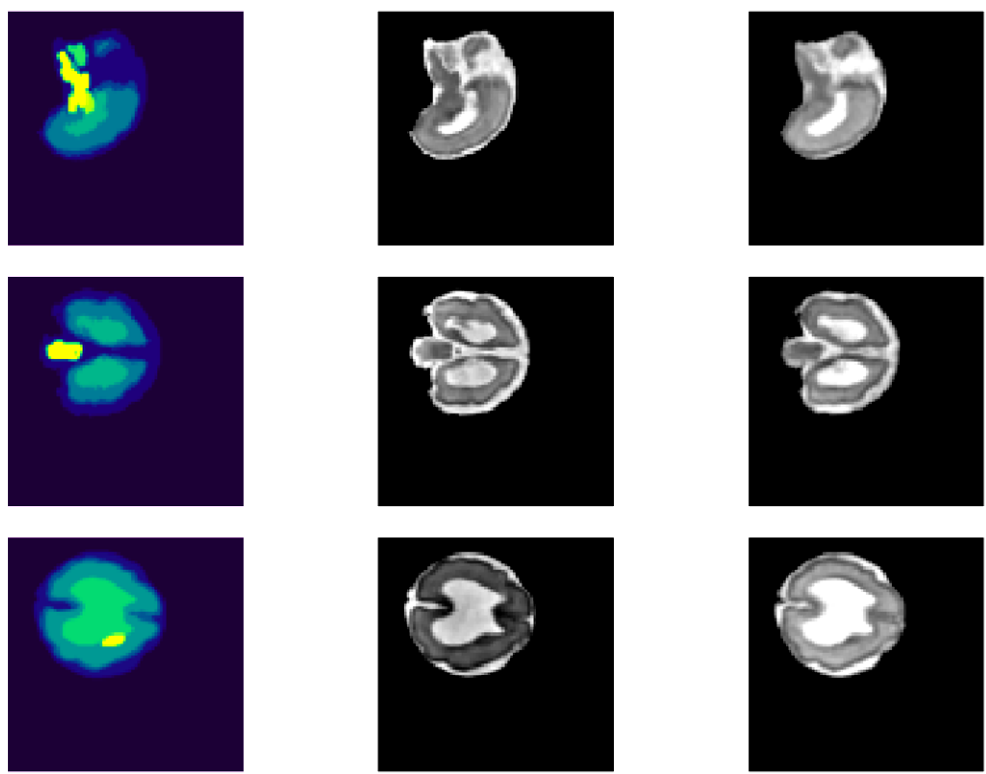

Figure A.3: Example of each of the SPADE networks trained on each orientation for only the 'mial' the training data. On the left you can see the input label, in the middle the original MRI scan, and on the right the synthesized image.

**A.3 Hydrocephalus degrees in coronal and axial orientation**

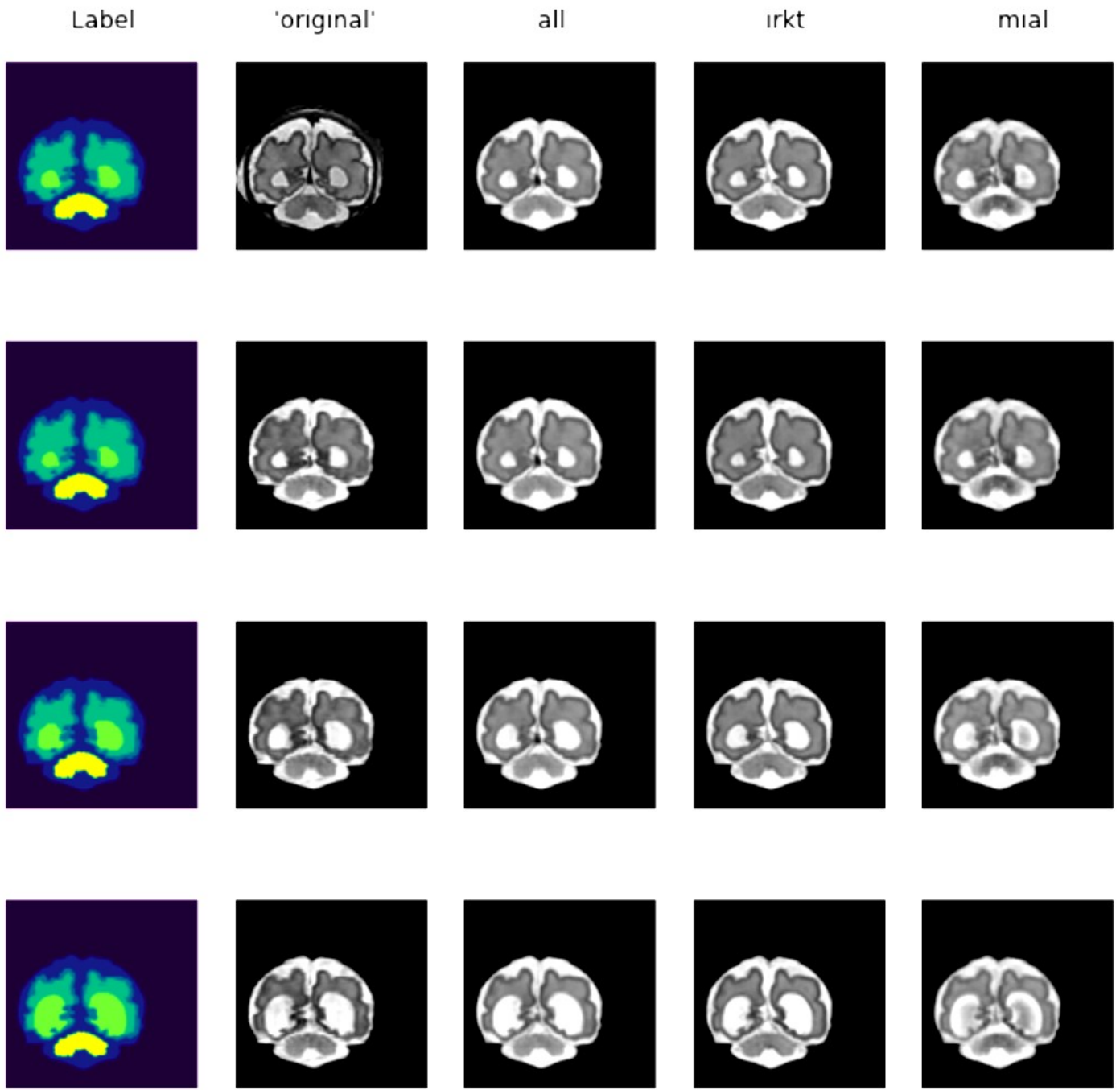

Figure A.4: Artificially created scenario with varying degrees of hydrocephalus. Coronal orientation.

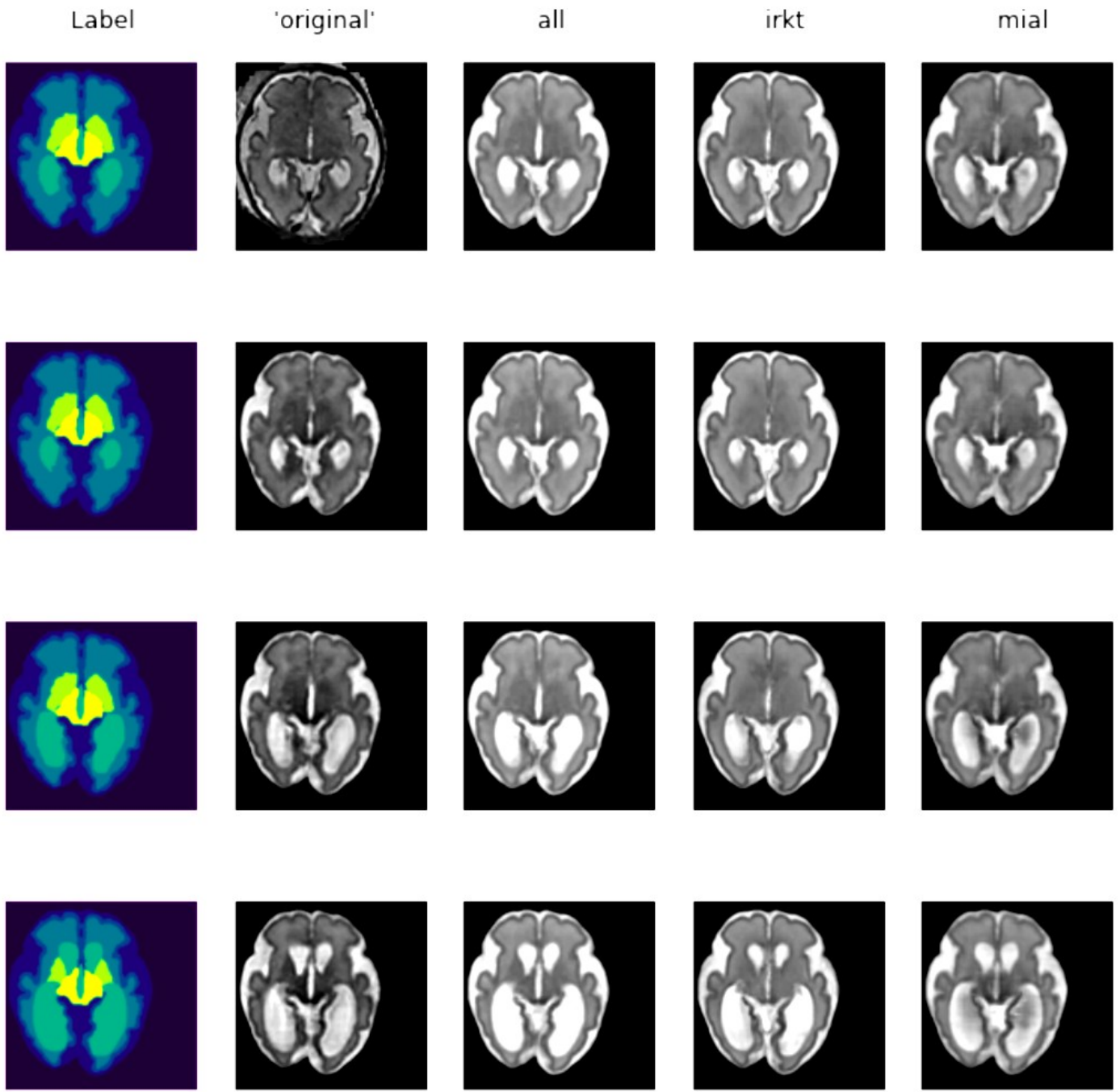

Figure A.5: Artificially created scenario with varying degrees of hydrocephalus. Axial orientation.